\documentclass[11pt]{llncs}
\begin{document}

\title{Quantum Digital Signature Based on Quantum One-way Functions
\thanks{This work was supported by the  Natural
Science Foundation of China under Grant No. 60273027 and the
National Grand Fundamental Research 973 Program of China under
Grant No. G1999035802}} 
\author{Xin L¨¹\"u,  Deng-Guo Feng
}
\institute{State Key Laboratory of Information Security,
\\Graduate School of Chinese Academy of Sciences,
 Beijing, 100039,China\\
 email: lx@is.ac.cn}

 \maketitle

\smallskip

\textbf{Abstract}: A quantum digital signature scheme  based on
quantum mechanics is proposed in this paper. The security of the
protocol relies on the existence of quantum one-way functions by
fundamental quantum principles. Our protocol involves a so-called
arbitrator who validates and authenticates the signed message.
This scheme uses public quantum keys to sign message and uses
quantum one-time pad to ensure the security of quantum information
on channel. To guarantee the authenticity of the transmitted
quantum states, a family of quantum stabilizer code is employed.
The proposed scheme presents a novel method to construct secure
quantum signature systems for future secure communications.

\smallskip

 \textbf{Key words}:
Digital signature; Quantum cryptography; Error correction code;
Quantum one-way functions

\section{Introduction}

Quantum cryptography aims at providing information security that
relies on the main properties of quantum mechanics. The most
successful topic of quantum  cryptography is quantum key
distribution (QKD), which was firstly invented by Bennett and
Brassard in 1984 \cite{BB84}. QKD is believed to be the first
practical quantum information processor and its unconditional
security has been proven \cite{Mayers01,Shor-simple}.

Other than QKD, quantum cryptography protocols are widely studied
in these years, such as quantum digital signature and quantum
message authentication. Digital signature is a main task in modern
cryptography and is widely used in today's communication systems.
Digital signature cares about the ``authenticity''  data on
channel \cite{Goldreich}. Informally, an unforgeable signature
scheme requires that each user be able to efficiently generate
his(her) own signature and  verify the validity of another user's
signature on a specific document, and no one be able to
efficiently generate the signatures of  other users to documents
that those users didn't sign.

Gottesman and Chuang proposed a quantum digital system \cite{QDS}
based on quantum mechanics, and claimed that the scheme was
absolutely secure, even against an adversary having unlimited
computational resources. The scheme, however, can only sign
classical bits string and can't deal with general quantum
superposition states. Zeng presented  an arbitrated quantum
signature scheme, the security of which is due to the correlation
of the GHZ triplet states and utilization of quantum one-time pad
\cite{Zeng}.In an arbitrated signature scheme, all communications
involve a so called arbitrator who has access to the contents of
the messages \cite{DS81}. The security of most arbitrated
signature schemes depends heavily on the trustworthiness of the
arbitrators. Zeng's protocol signs quantum messages which are
known to the signatory. It seems impossible to sign a general
unknown quantum state \cite{QDS,Zeng,Authentication}.

In this paper, we present a novel arbitrated quantum digital
signature scheme which can sign  general quantum states, the
security of which is based on a family of quantum one-way
functions by quantum information theory. This article is arranged
as below.

Section 2 introduces some definitions and preliminaries we will
use in the article. Section 3 describes the  proposed quantum
signature scheme. The security is considered in Section 4. Section
5 gives discussions and conclusions.

\section{Preliminaries}
\subsection{Quantum one-way function}
This section introduces a class of quantum one-way functions based
on the fundamental principles of quantum mechanics, which was
proposed by Gottesman and Chuang \cite{QDS} and the definitions
are presented as below.

\begin{definition}[quantum one-way function ] A function \begin{math}f:|x\rangle_{n_{1}}\mapsto|f(x)\rangle_{n_{2}} \end{math}
where \begin{math}x\in F_{2}^{n_{1}}\end{math} and
\begin{math}n_{1}\gg n_{2}\end{math}, is called a quantum one-way function
under physical mechanics if

(1) Easy to compute: There is a quantum polynomial-time algorithm
A such that on input \begin{math}|x\rangle\end{math} outputs
\begin{math}|f(x)\rangle\end{math}.

(2) Hard to invert: Given \begin{math}|f(x)\rangle\end{math}, it
is impossible to invert \textit{x}  by virtue of fundamental
quantum information theory.
 \end{definition}

What should be pointed out for the above definition is that the
condition \begin{math}n_{1}\gg n_{2}\end{math}  is necessary. By
Holevo's theorem \cite{Nielson}, no more than $n$ classical bits
of information can be obtained by measuring $n$ qubits quantum
states. Several means to construct quantum one-way function were
introduced by Gottesman and Chuang \cite{QDS} and here we choose
the quantum fingerprinting function \cite{finger} for the
candidate. The quantum fingerprinting function of a bit string
$u\in F_{2}^{w}$ is
\begin{equation}|f(u)\rangle=\frac{1}{\sqrt{m}}\sum_{l=1}^m(-1)^{E_{l}(u)}\cdot|l\rangle\end{equation}
where \begin{math}E:\{0,1\}^{w}\rightarrow\{0,1\}^{m}\end{math} is
a family of error correcting code with fixed
\begin{math}c>1,0<\delta<1\end{math} and
\begin{math}m=cw\end{math}. $E_{l}(u)$ denotes the $lth$ bit of $E(u)$. The distance between distinct code words $E(u_{1})$ and $E(u_{2})$ is at least
$(1-\delta)m$. Since two distinct code words can be equal in at
most $\delta m$ positions, for any $u_{1}\neq u_{2}$ we have
$\langle f(u_{1})|f(u_{2})\rangle\leq \delta m/m=\delta$. Here
\begin{math}f(u)\end{math} can be regarded as a class of quantum
one-way functions, which are easy to compute, but difficult to
reverse.

\subsection{Quantum stabilizer codes}
Quantum error correction code (QECC) is a way of encoding quantum
data (having $m$ qubits) into $n$ qubits (m$<$n), which protects
quantum states against the effects of noise. Quantum stabilizer
code is an important class of QECC and has been used to the other
subject of quantum information, such as quantum cryptography
\cite{Nielson}.

The Pauli operators $\{\pm I,\pm \sigma_{x},\pm \sigma_{y},\pm
\sigma_{z}\}$ constitute a group of order 8. The $n$-fold tensor
products of single qubit Pauli operators also form a group
$G_{n}=\pm\{I,\pm \sigma_{x},\pm \sigma_{y},\pm \sigma_{z}\}$, of
order $2^{2n+1}$. We refer to $G_{n}$ as the $n$-qubit Pauli
group. Let $S$ denote an abelian subgroup of the $n$-qubit Pauli
group $G_{n}$. Then the stabilizer codes $H_{S}\subseteq
H_{2^{2n}}$ satisfy,
\begin{equation}|\psi\rangle \in H_{S}, iff~~M|\psi\rangle=|\psi\rangle~~for~~all~~M\in S   \end{equation}

The group $S$ is called the stabilizer of the code, since it
preserves all of the codewords.

For stabilizer codes [[$n$, $k$, $d$]], the generators $M_{i }$
and the errors $E_{a}$, write

\begin{equation}M_{i}E_{a}=(-1)^{S_{ia}}E_{a}M_{i},i=1,\cdots,n-k
\end{equation}

The $s_{ia}'s$ constitute a syndrome for the error $E_{a}$, as
$(-1)^{S_{ia}}$ will be the result of measuring $M_{i }$ if the
error $E_{a }$ happens. For a nondegenerate code, $s_{ia}'s$ will
be distinct for all $E_{a}\in \varepsilon$, so that measuring the
$n-k$ stabilizer generators will diagnose the error completely.

\section{The Proposed Protocol}
\subsection{Security requirements}
The proposed scheme is a cryptographic protocol involving three
entities: a signatory Alice, a receiver Bob, and an arbitrator
Trent who authenticates and validates the signed message. The
security of the signature scheme depends much on the
trustworthiness of the arbitrator who has access to the contents
of the messages. The quantum digital signature discussed in this
article should meet the following security conditions:
\begin{enumerate}

\item Each user (Alice) can efficiently generate her own signature
on messages of his choice;

\item A receiver Bob can efficiently verify whether a given string
is a signature of another user's on specific message with Trent's
help;

\item The signatory can't disavow the message that she has signed;

\item It is infeasible to produce signatures of other users'
messages they haven't signed.
\end{enumerate}
\subsection{The  protocol}
\subsubsection{Key generation}
\begin{enumerate}

\item  Key distribution. Alice, Bob and Trent agree on some random
bits $K_{AT}$, $K_{AB}$ and $K_{TB}$ as their private keys.
$K_{AT}$  is shared between Alice and Trent, $K_{AB}$ is shared
between Alice and Bob and $K_{TB}$  between Trent and Bob .

$~~$ To ensure that the scheme is unconditionally secure, the keys
can be generated using quantum key distribution protocols, such as
BB84 or EPR protocol\cite{BB84,Nielson}.

\item Signature key generation. Alice generates 2\textit{k} random
secret strings
\begin{math}u_{i,j}\in F_{2}^{w}\end{math} and
computes
\begin{equation}|y_{i,j}\rangle=|f(u_{i,j})\rangle,1\leq i\leq 2n,j\in\{0,1\}\end{equation}
Here \begin{math}f:|x\rangle\mapsto|f(x)\rangle \end{math} is a
class of quantum one-way functions introduced in section 2. Alice
generates 4\textit{n} key pairs of
\begin{math}\{{u_{i,j},|y_{i,j} \rangle}\}_{j\in\{0,1\}}^{1\leq i\leq
2n}\end{math} and then publicly announces $\{|y_{i,j} \rangle\}_{j
\in\{0,1\}}^{1\leq i\leq 2n}$ as her public key and keeps
$\{u_{i,j} \}_{j \in\{0,1\}}^{1\leq i\leq 2n}$ as her private key.
\end{enumerate}

\subsubsection{Signature }

\begin{enumerate}

\item  Suppose Alice has a quantum state $|\psi\rangle \in
\mathcal{H}_{2^{n}}$ and wants to send it to Bob. Alice randomly
 selects bits strings $x \in F_{2}^{2n}$, $k$ for the stabilizer
codes $\{Q_{k}\}$ and $s$.  She $q$-encrypts $|\psi\rangle$ as
$\rho$ using $x$. Alice encodes $\rho$ according to $Q_{k}$ with
syndromes $s$ and obtains $\pi$.

\item Alice computes
\begin{equation}X=(x_{pre_{|s|}}\oplus
y)||(x_{suf_{2n-|s|}})\footnote{Suppose $s<2n$ in the algorithm.
Here, $x_{pre_{|y|}}$ denotes the first $|y|$ bits of $x$ and
$x_{suf_{2n-|y|}}$ denotes the last $2n-|y|$ bits of $x$, $a\oplus
b$ means the bit-by-bit XOR of the strings $a$ and $b$, namely
$a\oplus b=a_{1}\oplus b_{1},\cdots, a_{m}\oplus b_{m}$. The
symbol $``||''$ means concatenation of two binary
strings.}\end{equation} and generates four copies of $X's$
signature $|\Sigma_{K}(X)\rangle$ according to her key $K\in
\{u_{i,j},|y_{i,j}\rangle|1\leq i\leq 2n, j\in \{0,1\}\}$

\begin{equation}|\Sigma_{K}(X)\rangle=|y_{1,X_{1}}\otimes\ldots\otimes y_{2n,X_{2n}}\rangle=|a_{1}\otimes\dots\otimes a_{2n}\rangle=|a\rangle
\end{equation}

 Alice sends $\pi$ and two copies of $|\Sigma_{K}(X)\rangle$
to Bob. At the same time, she encrypts $\{s,k,x\}$ as $C_{1}$
using $K_{AT}$ \footnote{In this algorithm, we select classical
one-time-pad to encrypt classical message to ensure the
unconditional security.} and sends $C_{1}$ and two copies of
$|\Sigma_{K}(X)\rangle$ to Trent. We assume that each setting up
of a protocol has a unique sequence number.

\end{enumerate}

\subsubsection{Verification}
\begin{enumerate}

\item Trent receives $C'_{1}$ and two copies of
$|\Sigma'_{K}(X)\rangle=|a'\rangle$. Trent checks whether these
two copies of $|\Sigma'_{K}(x)\rangle$ he recieved are equivalent
by performing a quantum swap test circuit (QSTC \cite{finger}). If
any one of $|a'_{i}\rangle$'s fails the test, Trent aborts the
protocol. Trent decrypts $C'_{1}$ using his secure key $K_{AT}$
and obtains $\{s_{T},k_{T},x_{T}\}$. He computes
$|\Sigma_{K}(X)_{(T)}\rangle$ according to $x_{T}$ and Alice's
public keys. Trent compares
$|\Sigma_{K}(X)_{(T)}\rangle=|a\rangle_{T}$ to
$|\Sigma'_{K}(X)\rangle$. If any one of them fails the test, Trent
aborts the protocol. Trent encrypts $\{k_{T},x_{T}\}$ as $C_{2}$
using $K_{TB}$ and sends the ciphertext to Bob.

$~~$The comparison of two quantum states is less straightforward
than in the classical case because of the statistical properties
of quantum measurements. Another serious problem is that quantum
measurements usually introduce a noneligible disturbance of the
measured state. Here, we use the quantum swap test circuit (QSTC)
proposed in \cite{finger} to compare whether $|a_{i}\rangle_{T}$
and $|a'_{i}\rangle$ are equivalent or not. QSTC is a comparison
strategy with one-sided error probability $(1+\delta^{2}/2)$, and
each pair of the compared qubits has an inner product with an
absolute value at most $\delta$. Because there are $2n$ sets of
qubits to be compared, the error probability of the test can be
reduced to $(\frac{1+\delta^{2}}{2})^{2n}$, where $\langle
f_{i}|f_{j}\rangle\leq\delta$ with $i\neq j$, and $n$ is the
security parameter. Let the number of the incorrect keys be
$e_{j}$,  Bob  rejects it as invalid signature if $e_{j}>cM$. Here
$c$ is a threshold for rejection and acceptance in the protocol.

\item Bob has received Alice's information $[\pi'
,|\Sigma''_{K}(X)\rangle=|a''\rangle]$, $\pi'$ and  Trent's
message $C'_{2}$ now. He deciphers $C'_{2}$ as $\{k_{B},x_{B}\}$
and computes $X_{B}$ according to Eq.(5). He measures the syndrome
$s_{B}$ of the stabilizer code $Q_{k}$ on $\pi'$ and decodes the
qubits as $\rho'$. He encrypts $s_{B}$ as $C_{3}$ using parts of
$K_{TB}$ and sends it to Trent.

\item Trent encrypts $s_{T}$ as $C_{4}$ using parts of $K_{TB}$
and sends it to Bob.

\item Bob deciphers $C'_{4}$  and obtains $s_{T}$. He compares
$s_{B}$ to $s_{T}$ and aborts if any error is detected. Bob checks
whether these two copies of $|\Sigma''_{K}(X)\rangle$  are
equivalent by performing the QSTC. He computes quantum states
$|\Sigma(X)\rangle_{B}=|a\rangle_{B}$ using $X_{B}$ and Alice's
public keys $\{|y_{i,j}\rangle\}_{j \in\{0,1\}}^{1\leq i\leq 2n}$.
He verifies Alice's signature according to
\begin{equation}V_{K}(X_{B},|\Sigma'_{K}(X)\rangle)=True\Leftrightarrow \{|a'_{i}\rangle=|y_{i,X_{i}}\rangle=|a''_{i}\rangle_{B} \}_{1\leq i\leq 2n}\end{equation}

Bob $q$-decrypts $\rho'$ as $|\psi'\rangle$ according to $x_{B}$.

\end{enumerate}
\section{Security Analysis}
\subsection{Correctness}
\begin{theorem}[Correctness] Suppose all the entities involved in the
scheme follow the protocol, then Eq. (7) holds.
\end{theorem}
\textbf{Proof}. The correctness of the scheme can be seen by
inspection. In the absence of intervention, Trent will obtain
Alice's key $s,x,k$ and her signature of $X$. Trent verifies the
signature and sends $x,k$ secretly to Bob. Bob can successfully
decode and decipher the quantum states and verify Alice's
signature. Because Alice signs her message according to Eq. (6),
it's easy to verify that Eq. (7) holds.

\subsection{Security against repudiation}

Alice can't deny her signature. When  Alice disavows her
signature, Bob will resort to Trent. Bob sends one copy of the
signature $|\Sigma''_{K}(X)\rangle$ to Trent. Trent compares
$s_{B}$ and $|\Sigma''_{K}(X)\rangle$ with $s_{T}$ and his kept
copy of signature $|\Sigma'_{K}(X)\rangle$ Alice has sent to him.
If all these pass the test, Trent reveals that Alice is cheating
because $|\Sigma_{K}(X)\rangle$ contains Alice's signature on her
private keys $x$ and $s$. Otherwise, Trent concludes that the
signature has been forged by Bob or other attackers.

\subsection{Security against forgery}

\begin{theorem} Other entities forge Alice's signature with a
successful probability at most  $2^{-[(w-t\lceil
log_{2}m\rceil)+2n]}$.
\end{theorem}

\textbf{Proof.} Considering that an adversary (Eve or Bob)
controls the communication channels connecting Alice, Trent and
Bob and wants to forger Alice's signature. Here we present two
strategies that the attack Eve (Bob) can apply.
\begin{enumerate}

\item One is that she tries to alter the signed quantum states.
Eve intercepts $[\pi' ,|\Sigma'_{K}(X)\rangle]$. She keeps $\pi'$
and selects a random key  $x_{E}$ to encrypt another quantum
states $|\phi\rangle$ as $\tau$ and sends $[\tau
,|\Sigma'{K}(X)\rangle]$ to Bob. Because Eve knows nothing about
the stabilizer code $\{Q_{k}\}$ and syndrome $s$, her cheating
will be detected by Bob in the fourth step of the verification
phase when he compares  the syndrome $y$ to $y'$.

\item The second strategy is that the attacker tries to recover
Alice's private keys and generates a ``legal'' signature. Because
she knows nothing about Alice's private keys $x,y,k, K_{AT}$ and
$\{u_{i,j} \}_{j \in\{0,1\}}^{1\leq i\leq 2n}$. She can't compute
$x,y,k$ from the mixed state $\pi'$. According to  Holevo's
theorem \cite{Nielson}, Eve can obtain at most $t\lceil
log_{2}m\rceil$ bits of classical information about one of Alice's
signature key $\{u_{i,j} \}$ from Alice's public key. Here, $t$ is
a small natural number and we let $c=4$ in our scheme. Since she
lacks $w-t\lceil log_{2}m\rceil$ bits of information about any
private key which Alice hasn't revealed, she will only guess
correctly at most  $2^{-[w-t\lceil log_{2}m\rceil]}$ of it.
Therefore, the attacker can forger Alice's signature only with a
successful probability less than $2^{-[(w-t\lceil
log_{2}m\rceil)+2n]}$.

\end{enumerate}

\section{Concluding Remarks}
Designing quantum digital signature protocol is not trivial
because of several fundamental properties of quantum message.

The first and the most important property of quantum information
is the no-clone theorem, which forbids the unknown qubits
reproduction. For digital signature, how can we verify the
signature is indeed the signature on a specific state without
generating copies of the original message?

The second is the probability and irreversibility properties of
quantum measurement. That brings much troubles to decide whether a
state is a legal signature without changing that state.

The last property of secure quantum signature scheme is that it is
also a secure encryption scheme, which has been shown by Barnum
$et~al.$ in literature \cite{Authentication}.

In this article, we investigate how to span these obstacles and
present a quantum digital signature scheme. The security of the
scheme relies on the existence of a family of quantum one-way
functions by quantum principles.  The authenticity of the quantum
information is obtained by quantum error correction codes and
security of the quantum information on channel is ensured by
quantum one-time pad.

\end{document}